\documentclass[namedreferences]{SolarPhysics}
\usepackage[pdfborder={0 0 0 },urlcolor=blue]{hyperref}
\usepackage[optionalrh,solaenum]{spr-sola-addons} % For Solar Physics 
\usepackage{epsfig}                     % For eps figures, old commands
\usepackage{graphicx}                    % For eps figures, newer & more powerfull
\usepackage{psfrag}
\usepackage{color}                       % For color text: \color command
\usepackage{url}                         % For breaking URLs easily trough lines
\begin{document}

\begin{article}

\begin{opening}

\title{A Bayesian Analysis of the Correlations Among Sunspot Cycles}

%%%%%%%%%%%%%%%%%%%%%%%%%%%%%%%%%%%%%%%%%%%%%%%%%%%
%% Authors Names
%
\author{Y.~\surname{Yu}$^{1}$\sep
        D.A.~\surname{van Dyk}$^{2}$\sep
        V.L.~\surname{Kashyap}$^{3}$\sep
        C.A.~\surname{Young}$^{4}$ 
       }

%%%%%%%%%%%%%%%%%%%%%%%%%%%%%%%%%%%%%%%%%%%%%%%%%%%
%% Runningheads
%
\runningauthor{Y. Yu {\it et al.}}
\runningtitle{Bayesian Analysis of Sunspot Cycles}

%%%%%%%%%%%%%%%%%%%%%%%%%%%%%%%%%%%%%%%%%%%%%%%%%%%
%% Affilations 
%
  \institute{$^{1}$ University of California, Irvine, CA, USA 
                     email: \url{yamingy@uci.edu} \\ 
             $^{2}$ Imperial College London, London, UK 
                     email: \url{dvandyk@imperial.ac.uk} \\
             $^{3}$ Harvard-Smithsonian Center for Astrophysics, Cambridge, MA, USA 
                     email: \url{vkashyap@cfa.harvard.edu} \\
             $^{4}$ ADNET Systems Inc., NASA/GSFC, Greenbelt, MD, USA 
                     email: \url{c.alex.young@gsfc.nasa.gov}
             }

%%%%%%%%%%%%%%%%%%%%%%%%%%%%%%%%%%%%%%%%%%%%%%%%%%%
%%% Abstract 
\begin{abstract}
Sunspot numbers form a comprehensive, long-duration proxy of solar activity and have been used numerous times to empirically investigate the properties of the solar cycle.
%Numerous studies have discovered a number of correlations in the sunspot numbers over the 24 cycles that direct observational records are available.
A number of correlations have been discovered over the 24 cycles for which observational records are available.
Here we carry out a sophisticated statistical analysis of the sunspot record that reaffirms these correlations, and sets up an empirical predictive framework for future cycles.
An advantage of our approach is that it allows for rigorous assessment of both the statistical significance of various cycle features and the uncertainty associated with predictions.
We summarize the data into three {\sl sequential} relations that estimate the amplitude, duration, and time of rise to maximum for any cycle, given the values from the previous cycle.
We find that there is no indication of a persistence in predictive power beyond one cycle, and conclude that the dynamo does not retain memory beyond one cycle.
%Based on sunspot records up to Sep 2010, we obtain, for Cycle 24, an estimated maximum smoothed monthly sunspot number of 78 $\pm$ 18, to occur in April/May 2014 $\pm$ 8 months.
Based on sunspot records up to October 2011, we obtain, for Cycle 24, an estimated maximum smoothed monthly sunspot number of 97 $\pm$ 15, to occur in January--February 2014 $\pm$ 6 months.
\end{abstract}

%%%%%%%%%%%%%%%%%%%%%%%%%%%%%%%%%%%%%%%%%%%%%%%%%%%
%% Keywords
%
%\keywords{}

\end{opening}

\begin{section}{Introduction}
\label{sec:intro}

Sunspot numbers constitute the longest continuous record of observations in astronomy, having been recorded in observatories worldwide since the Dalton minimum, and they are available as monthly estimates since 1749. 
Because of the multi-generational span, both the quality of the observations and the techniques used to record them have varied, and thus these data present a challenge for interpretation.
The data are maintained at the Solar Influences Data Analysis Center in Belgium (\url{http://sidc.oma.be}) as the international sunspot numbers (SSNs) and their suitability as a proxy for solar activity has been found to be good by comparison with other proxies (Waldmeier, 1971).
(Recent efforts to recalibrate the SSN have suggested potential differences between the different proxies ({\it e.g.} Svalgaard, 2010).
However, we do not include these corrections in our analysis because first, there is a possibility that a fundamental change has occurred in the manifestation of the activity proxies in recent years, and second, there is no evidence that the calibration needs to be changed over the entire historical record.)
The long duration of the dataset allows us to analyze a much larger number of cycles than with other, more recently developed proxies of solar activity (such as sunspot area, umbral fields, 10.7~cm flux, {\it etc.}).

Accurate prediction of solar activity cycles is an important area of research since variations in ``space weather" caused by solar activity affect radio communication, the performance of low-Earth orbit satellites, and geomagnetic activity ({\it e.g.} aurorae).
Studying the behavior of sunspot cycles is important not only for understanding the physics of solar activity, but also for the planning of space missions. 
Sunspot numbers are valuable as an indicator of solar activity, and identifying recurring patterns in sunspot cycles is therefore
crucial from both an empirical and a theoretical perspective.
Indeed, the 11-year activity cycle of the Sun was first discovered by noticing the same cycle in SSNs (Wolf, 1852).
While correlations with solar activity have been identified in other indicators (solar flare numbers, sunspot areas, 10.7cm flux, {\it etc.}; see Hudson, 2007) and studies of solar activity have been extended back many millenia using dendrochronological data (Bonev {\it et al.}, 2003; Solanki {\it et al.}, 2004), the SSN data are the first rung in the ladder to calibrate all observations of the solar activity cycle.
Waldmeier (1935, 1939) noted that sunspot cycles tend to take less time to rise to maximum than to fall to minimum.
Other important relations, such as the correlation between the duration of a cycle and the amplitude of the next cycle (amplitude--period effect), can be used to predict characteristics of the upcoming cycle years in advance ({\it e.g.} Hathaway {\it et al.}, 1994, 2002; Watari, 2009).
Here we analyze the SSN data to derive statistically meaningful phenomenological correlations between the various parameters that observationally define a solar cycle.
Such correlations act as constraints on theoretical models of dynamo action that seek to explain solar activity (Sch\"{u}ssler, 2007).
Analysis of long-duration activity cycles (Usoskin {\it et al.}, 2007) suggest that they are driven by a stochastic or chaotic process, and it is therefore necessary to identify the statistical properties of the activity cycles in the current era.

Prediction methods for upcoming cycles include those based on i) solar dynamo models (Choudhuri, 1992; Charbonneau and Dikpati, 2000; Dikpati {\it et al.}, 2006; Choudhuri {\it et al.}, 2007; Charbonneau, 2007), ii) precursors such as geomagnetic {\it aa} indices (Hathaway and Wilson, 2006), and iii) statistical analyses with historical data (Hathaway {\it et al.}, 1994; Benestad, 2005; Xu {\it et al.}, 2008; Gil-Alana, 2009), among others.
Pesnell (2008) gives a review of a large range of predictions made for the upcoming Cycle 24. 
%Numerous studies have analyzed solar activity from a statistical perspective.  
Given the current debate over the amplitude of Cycle 24, for which different physical models yield substantially different predictions (see, for example, Dikpati and Gilman, 2006, and Choudhuri {\it et al.}, 2007), a statistical method that uses only the SSN data will provide a useful benchmark for comparison.
Methods based on statistical extrapolation, however, rely on various assumptions just as physical models do, and may not fully account for the uncertainties involved.
For example, for Cycle 23, the smoothed maximum sunspot number as predicted by several researchers varied considerably, from 80 to 210.
Kane (2001) noted that among 20 predictions, only 8 were within a reasonable range of the actual value. 

%Hathaway et al.\ (1994) advocate a technique that represents each cycle as a curve with four parameters: starting time, %amplitude, rise time, and asymmetry.
%Relations between these cycle parameters, such as a weak correlation between %amplitude and length of the previous cycle, are found, and are employed for prediction. 

We describe our analysis of SSN data in Section~\ref{sec:analysis}, and in particular describe the statistical model used to determine the correlations between the parameters defining a given cycle from the previous cycle (Section~\ref{sec:statmodel}) and fitting the model to the data (Section~\ref{sec:fitting}).
We report the resulting correlations and discuss their implications in Section~\ref{sec:discuss}, and summarize our work in Section~\ref{sec:summary}.

\end{section}

\begin{section}{Analysis}
\label{sec:analysis}

\subsection{Two-Stage Statistical Model}
\label{sec:statmodel}

This article proposes a two-stage statistical model that accounts for the uncertainty both in smoothed monthly sunspot numbers and in predicting future cycle characteristics from historical data.
In the first stage of the statistical model, cycle characteristics, such as amplitude, duration, and rising time, are estimated from raw SSNs.
Then, in the second stage, relations between the characteristics of consecutive cycles are examined.
This results in three sequential relations that summarize known features of sunspot cycles such as the Waldmeier effect and the correlation between amplitudes of successive cycles.
These statistical properties place a constraint on any physical model that attempts to explain the solar cycle behavior. 

In this section we fit the two stages of the statistical model separately. This involves first modeling the cycles
in Section~\ref{sec:stage1} and then modeling the relationships among the parameters that
describe the cycles in Section~\ref{sec:stage2}.  We discuss how these two stages can be combined
into a single coherent statistical model in Section~\ref{sec:fitting} and show the results of this
coherent fit in Section~\ref{sec:discuss}. 

\subsubsection{Stage One: Modeling the Cycles}
\label{sec:stage1}

When extracting cycle length, rising time, and amplitude information from SSN data, we adopt an approach similar to the two-parameter curve fitting of Hathaway {\it et al.} (1994; see also Sabarinath and Anilkumar, 2008, and Volobuev, 2009, who propose other functional forms).
For cycle $i$, suppose $t_0^{(i)}$ is the starting time, $t_{\rm max}^{(i)}$ is the time of the cycle maximum, $t_1^{(i)}$ is the end time, $c_i$ is the amplitude, and $U_t$ is a parameter that captures the ``average solar activity level'' at time $t$.
We postulate that 

\begin{itemize}
\item
for the rising phase $t<t_{\rm max}^{(i)}$
\begin{equation}
\label{rise}
U_t=c_i\left(1-\left(\frac{t_{\rm max}^{(i)}-t}{t_{\rm max}^{(i)}-t_0^{(i)}}\right)^{\alpha_1}\right)\quad {\rm and}
\end{equation}
\item
for the declining phase $t>t_{\rm max}^{(i)}$
\begin{equation}
U_t=c_i\left(1-\left(\frac{t-t_{\rm max}^{(i)}}{t_1^{(i)}-t_{\rm max}^{(i)}}\right)^{\alpha_2}\right),
\label{fall}
\end{equation}
\end{itemize}
where $\alpha_1, \alpha_2>1$ are shape parameters assumed to be the same for all cycles.
%The vector $\alpha=(\alpha_1, \alpha_2)$ is assumed to be the same for all cycles.
A curve described by the two postulates is illustrated in Figure~\ref{fig:cycleparam}.
We do not assume that the starting point of the next cycle [$t_0^{(i+1)}$] is identical to the end point of the current cycle, $t_1^{(i)}$.
When $t_0^{(i+1)}<t_1^{(i)}$, the two cycles overlap, and the activity level [$U_t$] during the overlapping period is defined as the sum of the contributions of the form (\ref{fall}) and (\ref{rise}) from these two cycles.
We adopt this parameterization because it is simple, flexible, easy to interpret, and fits the data well. 

\begin{figure}[htb!]
\caption{Parameterized form of a solar cycle.  We illustrate $U_t^2$ with $c_i=10,\ \alpha_1=1.9$ and $\alpha_2=1.1$, where $U_t$ is specified by (\ref{rise}) and (\ref{fall}).  Because we model after a square-root transformation, $U_t^2$ approximates the shape of a cycle on the original scale of the SSN data.}
\label{fig:cycleparam}
\begin{center}
\psfrag{a}{$t_0^{(i)}$}
\psfrag{b}{$t_{\rm max}^{(i)}$}
\psfrag{c}{$t_1^{(i)}$}
\psfrag{Ut}{$U_t^2$}
\psfrag{year}{$t$}
\includegraphics[width=2.5in,angle=270]{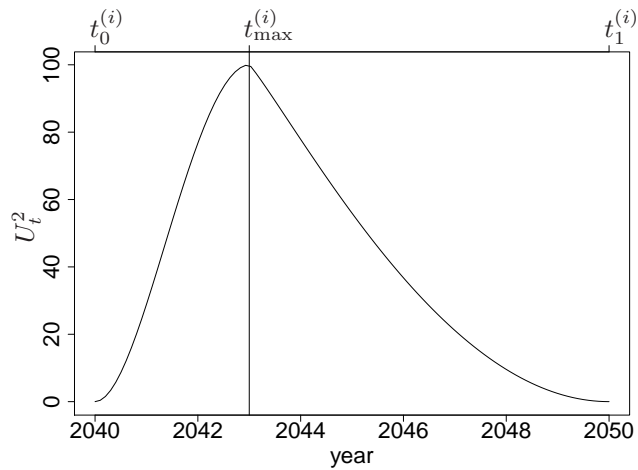}
\end{center}
\end{figure}

Given a total of $k=25$ cycles (including the incomplete Cycles 0 and 24), we relate the parameters $\alpha=(\alpha_1, \alpha_2)$ as well as $T_0=(t_0^{(i)},\ i=0, \ldots, k-1)$, $T_{\rm max}=(t_{\rm max}^{(i)},\ i=0,\ldots, k-1)$, $T_1=(t_1^{(i)},\ i=0, \ldots, k-1)$, and $C=(c_i,\ i=0, \ldots, k-1)$, to the observed data by a linear model: 
\begin{equation}
\label{eq:stage1}
\sqrt{Y_t}|(T_0, T_{\rm max}, T_1, C, \alpha, \beta, \sigma^2)\stackrel{\rm ind}{\sim} N(\beta +U_t,\ \sigma^2),
\end{equation}
where the parameter $\beta$ may be regarded as a baseline, and we model $Y_t$, the monthly average sunspot number at time $t$, after a square-root transformation.
This transformation is used to stabilize the variance, since higher sunspot numbers are also associated with higher variability. 
(A variance-stabilizing transformation is a mapping [$f(y)$] of the data [$y$] such that the variability of $f(y)$ is constant relative to its mean value.  For example, for counts data that follow a ${\rm Poisson}(\mu)$ distribution, the variance increases with the mean $\mu$.  The variance of $\sqrt{y}$, however, is approximately the same for different $\mu$.  The situation with SSNs is similar and a square-root transformation is suitable.  Note that results such as predictions are easily obtained on the original scale by inverting the transformation.)

Our approach differs from other curve-fitting methods ({\it e.g.} Hathaway {\it et al.}, 1994) in that we model all cycles jointly and we estimate the starting and ending points of the cycles from the data rather than fixing them in advance. 
In addition, we model the monthly sunspot numbers directly rather than smooth them beforehand, {\it e.g.} by a moving average, as is often done in other statistically based prediction methods. 
 It may be interesting to analyze daily data in a future work; see Noble and Wheatland (2012) for an analysis
of the daily fluctuations.  One intrinsic difficulty with daily data, however, is that the same spot or spot 
group is counted every day until it disappears or rotates away over the limb (aside from the issue of merging 
or splitting spots).  We choose the monthly data partly to alleviate this problem.

%The monthly sunspot numbers are approximately independently distributed (apart from a few long lasting spots that persist for more than one rotation), and smoothing them tends to underestimate the uncertainty in the overall analysis.  

When $T_0$ and $T_{\rm max}$ are treated as unknown parameters, model (\ref{eq:stage1}) is not the usual linear regression model that can be easily fit by ordinary least squares.
To overcome such difficulties, we adopt a Bayesian approach and use Markov chain Monte Carlo (MCMC) to simulate samples from the {\it posterior} distribution.
We discuss Bayesian analysis, MCMC, and how we use them to fit our model in Section~\ref{sec:fitting}.
The fitted model and residuals are illustrated in Figure~\ref{fig:ssn}.
These are based on a random draw from the posterior distribution after fitting (\ref{eq:stage1}).
Plots using multiple posterior draws are qualitatively the same and are omitted.
The residuals are defined as $\sqrt{Y_t} - \beta - U_t$ as in (\ref{eq:stage1}).
The residual plot reveals a reasonably good fit, given the simple functional form of (\ref{rise}) and (\ref{fall}).
%The fit is slightly better for recent data (${\rm year}>1850$) than for data in the more distant past. 

\begin{figure}[htb!]
\caption{Fitted values and residuals for the regression model in (\ref{eq:stage1}) for one posterior sample.
Top: sunspot numbers with fitted values of $T_{\rm max}$ represented as vertical lines (only the last few cycles are shown for illustration). 
Bottom: histogram of residuals with approximating normal density curve.
}
\label{fig:ssn}
\begin{center}
  \psfrag{histogram of residuals}{Residuals: $\sqrt{Y_t} - \beta - U_t$}
  \includegraphics[width=3.6in, angle=270]{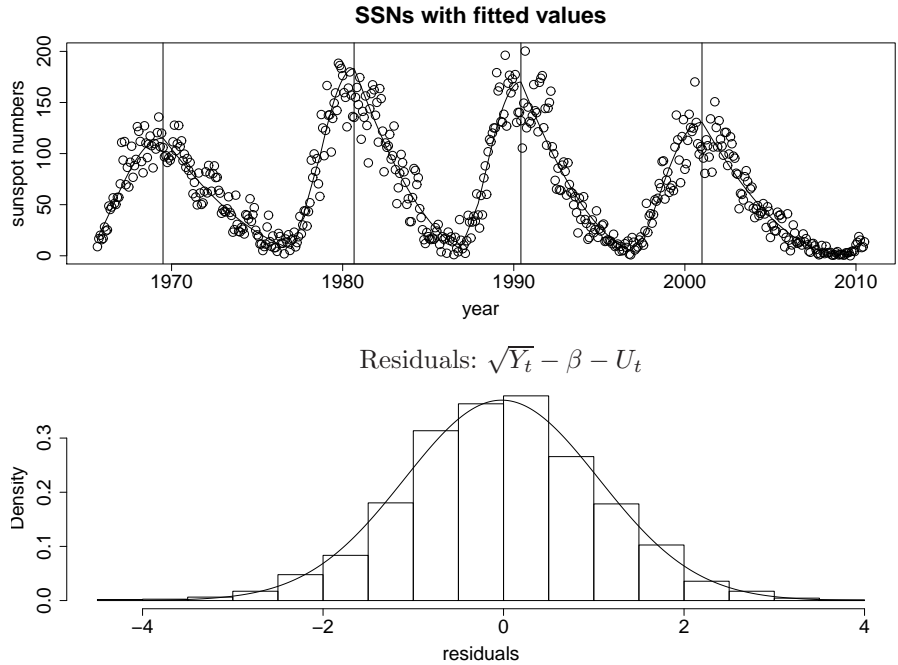}
\end{center}
\end{figure}

Note that (\ref{eq:stage1}) alone does not specify relationships between consecutive cycles, and is therefore not useful for predicting cycle characteristics of entirely new cycles.
In Section~\ref{sec:stage2}, we explore relationships between cycle characteristics, and build additional structure on Equation (\ref{eq:stage1}) to enhance its usefulness in predictions. 

Table~\ref{tbl:common:par} displays the fitted values and standard errors for some parameters that are common to all cycles.  The difference between $\alpha_1$ and $\alpha_2$ suggests an asymmetry in shape between the rising and declining phases of a cycle.  We have also fitted with individual $(\alpha_1, \alpha_2)$ for each cycle, but the results are similar and hence omitted. 
\begin{center}
\begin{table}[htb!]
\caption{Fitted values (posterior means) and standard errors (posterior standard deviations) for $\alpha_1,\ \alpha_2,\ \beta$ and $\sigma^2$.}
\label{tbl:common:par}
\begin{tabular}{rrrr}
\hline
 $\alpha_1$  & $\alpha_2$ & $\beta$ & $\sigma^2$  \\    
\hline
  $1.9\pm 0.11$       &     $1.1\pm 0.04$    &    $0.47\pm 0.18$  & $1.2\pm 0.03$ \\
\hline 
\end{tabular}
\end{table}
\end{center}
Table~\ref{tbl:fitparam} summarizes the parametric fits to each cycle.  (Again, the model-fitting procedure will be described in Section~\ref{sec:fitting}.) Note that the interval between when the old cycle ends and the new cycle begins is usually negative, suggesting that the new cycle begins before the old cycle ends.
This is consistent with recent results based on torsional oscillations (Hill {\it et al.}, 2010). 

\begin{table}[htb!]
\caption{Cycle-profile parameter posterior means for each cycle}
\label{tbl:fitparam}
\begin{tabular}{crrrrrr}     
\hline
cycle & amplitude & rise [yr] & fall [yr] & gap [yr] & start [yr] & end [yr]\\
\hfil & $c_i \sim \sqrt{{\rm SSN}}$ & $t_{\rm max}-t_0$ & $t_1-t_{\rm max}$ & $t_0^{i+1}-t_1^{i}$ & $t_0$ & $t_1$\\
\hline
 0 & $ 9.05{_{-0.34}^{+ 0.36}}$ & $ 5.00{_{-2.83}^{+ 3.33}}$ & $ 6.43{_{-0.35}^{+ 0.40}}$ & $-1.40{_{-0.18}^{+ 0.15}}$ 
 & $1745.3{_{-3.2}^{+ 2.6}}$ & $1756.6{_{-0.2}^{+ 0.2}}$ \\
 1 & $ 8.55{_{-0.22}^{+ 0.23}}$ & $ 6.41{_{-0.32}^{+ 0.19}}$ & $ 7.23{_{-0.40}^{+ 0.44}}$ & $-2.10{_{-0.24}^{+ 0.26}}$ 
 & $1755.2{_{-0.2}^{+ 0.2}}$ & $1768.9{_{-0.3}^{+ 0.3}}$\\
 2 & $10.32{_{-0.21}^{+ 0.23}}$ & $ 3.41{_{-0.25}^{+ 0.17}}$ & $ 7.38{_{-0.31}^{+ 0.37}}$ & $-1.10{_{-0.24}^{+ 0.18}}$ 
 & $1766.8{_{-0.1}^{+ 0.1}}$ & $1777.5{_{-0.3}^{+ 0.3}}$\\
 3 & $12.25{_{-0.24}^{+ 0.25}}$ & $ 2.04{_{-0.21}^{+ 0.21}}$ & $ 7.49{_{-0.24}^{+ 0.26}}$ & $-1.07{_{-0.18}^{+ 0.16}}$ 
 & $1776.4{_{-0.1}^{+ 0.1}}$ & $1786.0{_{-0.2}^{+ 0.2}}$\\
 4 & $11.57{_{-0.22}^{+ 0.24}}$ & $ 3.07{_{-0.24}^{+ 0.26}}$ & $11.73{_{-0.24}^{+ 0.27}}$ & $-1.15{_{-0.18}^{+ 0.24}}$ 
 & $1784.9{_{-0.2}^{+ 0.2}}$ & $1799.7{_{-0.2}^{+ 0.2}}$\\
 5 & $ 6.98{_{-0.24}^{+ 0.29}}$ & $ 5.99{_{-0.32}^{+ 0.34}}$ & $ 5.59{_{-0.42}^{+ 0.42}}$ & $ 1.52{_{-0.44}^{+ 0.48}}$ 
 & $1798.6{_{-0.1}^{+ 0.1}}$ & $1810.1{_{-0.2}^{+ 0.2}}$\\
 6 & $ 6.20{_{-0.22}^{+ 0.25}}$ & $ 5.93{_{-0.45}^{+ 0.40}}$ & $ 5.88{_{-0.38}^{+ 0.37}}$ & $ 0.07{_{-0.32}^{+ 0.26}}$ 
 & $1811.7{_{-0.2}^{+ 0.3}}$ & $1823.5{_{-0.2}^{+ 0.3}}$\\
 7 & $ 8.33{_{-0.22}^{+ 0.24}}$ & $ 6.82{_{-0.23}^{+ 0.18}}$ & $ 4.73{_{-0.24}^{+ 0.36}}$ & $-0.73{_{-0.19}^{+ 0.14}}$ 
 & $1823.5{_{-0.1}^{+ 0.1}}$ & $1835.1{_{-0.2}^{+ 0.2}}$\\
 8 & $11.63{_{-0.25}^{+ 0.27}}$ & $ 2.81{_{-0.16}^{+ 0.19}}$ & $ 8.43{_{-0.35}^{+ 0.32}}$ & $-1.67{_{-0.16}^{+ 0.18}}$ 
 & $1834.4{_{-0.1}^{+ 0.1}}$ & $1845.6{_{-0.3}^{+ 0.2}}$\\
 9 & $10.48{_{-0.24}^{+ 0.23}}$ & $ 4.92{_{-0.25}^{+ 0.25}}$ & $ 8.71{_{-0.30}^{+ 0.37}}$ & $-1.00{_{-0.17}^{+ 0.25}}$ 
 & $1843.9{_{-0.2}^{+ 0.2}}$ & $1857.6{_{-0.2}^{+ 0.3}}$\\
10 & $ 9.81{_{-0.21}^{+ 0.25}}$ & $ 3.76{_{-0.26}^{+ 0.24}}$ & $ 9.16{_{-0.41}^{+ 0.34}}$ & $-1.79{_{-0.21}^{+ 0.20}}$ 
& $1856.6{_{-0.2}^{+ 0.1}}$ & $1869.5{_{-0.2}^{+ 0.3}}$\\
11 & $11.05{_{-0.26}^{+ 0.25}}$ & $ 3.20{_{-0.20}^{+ 0.13}}$ & $ 8.07{_{-0.24}^{+ 0.26}}$ & $-0.28{_{-0.22}^{+ 0.19}}$ 
& $1867.7{_{-0.1}^{+ 0.1}}$ & $1879.0{_{-0.1}^{+ 0.2}}$\\
12 & $ 8.23{_{-0.22}^{+ 0.24}}$ & $ 4.72{_{-0.47}^{+ 0.44}}$ & $ 7.29{_{-0.54}^{+ 0.46}}$ & $-0.60{_{-0.23}^{+ 0.27}}$ 
& $1878.7{_{-0.1}^{+ 0.1}}$ & $1890.7{_{-0.3}^{+ 0.3}}$\\
13 & $ 9.15{_{-0.23}^{+ 0.23}}$ & $ 3.47{_{-0.30}^{+ 0.28}}$ & $ 8.99{_{-0.32}^{+ 0.34}}$ & $-0.65{_{-0.19}^{+ 0.23}}$ 
& $1890.1{_{-0.1}^{+ 0.1}}$ & $1902.6{_{-0.2}^{+ 0.2}}$\\
14 & $ 8.11{_{-0.25}^{+ 0.26}}$ & $ 4.98{_{-0.40}^{+ 0.27}}$ & $ 6.77{_{-0.45}^{+ 0.48}}$ & $-0.01{_{-0.24}^{+ 0.26}}$ 
& $1901.9{_{-0.2}^{+ 0.2}}$ & $1913.7{_{-0.3}^{+ 0.2}}$\\
15 & $ 9.44{_{-0.23}^{+ 0.26}}$ & $ 4.45{_{-0.28}^{+ 0.22}}$ & $ 6.27{_{-0.27}^{+ 0.39}}$ & $-0.95{_{-0.21}^{+ 0.12}}$ 
& $1913.7{_{-0.2}^{+ 0.1}}$ & $1924.4{_{-0.2}^{+ 0.3}}$\\
16 & $ 8.82{_{-0.25}^{+ 0.29}}$ & $ 4.61{_{-0.36}^{+ 0.39}}$ & $ 6.79{_{-0.38}^{+ 0.46}}$ & $-0.77{_{-0.23}^{+ 0.19}}$ 
& $1923.4{_{-0.2}^{+ 0.2}}$ & $1934.8{_{-0.2}^{+ 0.2}}$\\
17 & $10.70{_{-0.23}^{+ 0.24}}$ & $ 4.25{_{-0.17}^{+ 0.25}}$ & $ 7.43{_{-0.35}^{+ 0.32}}$ & $-1.27{_{-0.23}^{+ 0.18}}$ 
& $1934.0{_{-0.1}^{+ 0.1}}$ & $1945.7{_{-0.2}^{+ 0.2}}$\\
18 & $12.49{_{-0.28}^{+ 0.27}}$ & $ 4.02{_{-0.19}^{+ 0.23}}$ & $ 6.73{_{-0.23}^{+ 0.19}}$ & $-0.51{_{-0.08}^{+ 0.17}}$ 
& $1944.5{_{-0.1}^{+ 0.1}}$ & $1955.2{_{-0.1}^{+ 0.1}}$\\
19 & $14.19{_{-0.26}^{+ 0.29}}$ & $ 3.23{_{-0.15}^{+ 0.10}}$ & $ 7.81{_{-0.16}^{+ 0.19}}$ & $-1.04{_{-0.13}^{+ 0.12}}$ 
& $1954.7{_{-0.1}^{+ 0.1}}$ & $1965.8{_{-0.2}^{+ 0.2}}$\\
20 & $10.61{_{-0.24}^{+ 0.26}}$ & $ 4.38{_{-0.30}^{+ 0.29}}$ & $ 9.97{_{-0.47}^{+ 0.53}}$ & $-2.06{_{-0.29}^{+ 0.31}}$ 
& $1964.7{_{-0.1}^{+ 0.1}}$ & $1979.1{_{-0.4}^{+ 0.3}}$\\
21 & $13.04{_{-0.29}^{+ 0.25}}$ & $ 3.46{_{-0.21}^{+ 0.21}}$ & $ 7.32{_{-0.32}^{+ 0.26}}$ & $-1.01{_{-0.16}^{+ 0.09}}$ 
& $1977.0{_{-0.1}^{+ 0.1}}$ & $1987.8{_{-0.2}^{+ 0.2}}$\\
22 & $12.91{_{-0.24}^{+ 0.29}}$ & $ 3.42{_{-0.25}^{+ 0.18}}$ & $ 7.19{_{-0.28}^{+ 0.22}}$ & $-0.88{_{-0.12}^{+ 0.13}}$ 
& $1986.8{_{-0.1}^{+ 0.1}}$ & $1997.4{_{-0.1}^{+ 0.2}}$\\
23 & $10.90{_{-0.26}^{+ 0.26}}$ & $ 4.40{_{-0.23}^{+ 0.27}}$ & $ 8.31{_{-0.39}^{+ 0.36}}$ & $-0.26{_{-0.33}^{+ 0.34}}$ 
& $1996.5{_{-0.2}^{+ 0.2}}$ & $2009.2{_{-0.2}^{+ 0.2}}$\\
\hline
\end{tabular}
\end{table}

\subsubsection{Stage Two: Modeling Relationships Between Consecutive Cycles}
\label{sec:stage2}

The goal at this stage of the statistical model is to build up an empirical mechanism that generates the amplitude, duration, and rising time of a cycle, given those of the previous cycle.  As a preliminary check, we carry out correlations between different parameters that define the model in a given cycle as well as between adjacent cycles.  The results of the correlation analyses are reported in Tables~\ref{tbl:corrcyc}-\ref{tbl:corrback}.  The tables list Spearman's rank coefficient [$\rho$] (Kendall, 1975) computed using the best-fit parameter values and the corresponding $p$-value. 
(The error bars on $\rho$ are computed from 200 posterior samples of the parameters and represent the robustness of the correlation.
The $p$-value represents the probability that such a correlation may be obtained by chance.)  Within a cycle (Table~\ref{tbl:corrcyc}),
a clear anti-correlation is seen between the amplitude [$c_i$] and the rise time [$t_{\rm max}^i-t_0^i$], as well as between rise time and fall time [$t_1^i-t_{\rm max}^i$]; {\it i.e.} strong cycles rise to maximum quickly (Hathaway {\it et al.}, 1994), and when they rise quickly they tend to decline slowly.  Correlations also appear to be present between amplitude and fall time, with strong cycles correlated with long declines (consistent with the above anticorrelations), and inversely between fall time and cycle gap [$t_0^{i+1}-t_1^i$].
 %suggesting that the rise time is of more importance to the duration than the fall time.
However, the direct correlation between the rise time and cycle gap is not statistically significant.

We show the correlations of a parameter with the parameter values of the following cycle in Table~\ref{tbl:corrforw} and with the parameter values in the preceding cycle in Table~\ref{tbl:corrback}.  Only the amplitude is significantly correlated across the cycles.
There is weak evidence for a correlation between the current cycle amplitude and the fall time of the next cycle, and inversely between the previous cycle gap and the current amplitude.  For more recent determinations of correlations across cycles, see Kane (2008), Kakad (2011), Ramesh and 
Lakshmi (2012); see also Vaquero and Trigo (2008) for a cautionary note. 

\begin{table}[htb!]
\caption{Correlations within cycle}
\label{tbl:corrcyc}
\begin{tabular}{llll}     
\hline
\hfil &  risetime  &  falltime  &  gap \\
\hfil &  $t_{\rm max}^i-t_0^i$  &  $t_1^i-t_{\rm max}^i$  &  $t_0^{i+1}-t_1^i$ \\
\hline
amplitude  &  $\rho=-0.81~\pm 0.06$ &  $\rho= 0.48~\pm 0.07$ &  $\rho=-0.32~\pm 0.06$ \\
 $c_i$  &  $p= 0.00$ &  $p= 0.02$ &  $p= 0.13$ \\
\hline
risetime  &  \hfil &  $\rho=-0.62~\pm 0.09$ &  $\rho= 0.29~\pm 0.09$ \\
 $t_{\rm max}^i-t_0^i$  &  \hfil &  $p= 0.00$ &  $p= 0.18$ \\
\hline
falltime  &  \hfil &  \hfil &  $\rho=-0.46~\pm 0.08$ \\
 $t_1^{i}-t_{\rm max}^i$  &  \hfil &  \hfil &  $p= 0.02$ \\
\hline
\end{tabular}
\end{table}

%\begin{table}
%\caption{Correlations with adjacent cycle}
%\label{tbl:corradj}
%\begin{tabular}{lc|l|l|l|l}     
%\hline
%\hfil & \hfil & amplitude$+$ & risetime$+$ & falltime$+$ & interval$+$ \\
%\hline
%amplitude$-$ & $\rho$ & $0.46 \pm 0.02$ &  $-0.21 \pm 0.06$ &  $0.39 \pm 0.05$ &  $-0.06 \pm 0.07$ \\
%\hfil &  $p$ & $0.03$ &  $0.34$ &  $0.07$ &  $0.80$ \\
%\hline
%risetime$-$ & $\rho$ &  $-0.17 \pm 0.08$ &  $0.05 \pm 0.10$ &  $-0.27 \pm 0.07$ &  $-0.06 \pm 0.13$ \\
%\hfil &  $p$ & $ 0.44$ &  $0.83$ &  $0.21$ &  $0.37$ \\
%\hline
%falltime$-$ &  $\rho$ & $ 0.07 \pm 0.05$ &  $-0.08 \pm 0.06$ &  $0.20 \pm 0.07$ &  $0.19 \pm 0.07$ \\
%\hfil &  $p$ & $ 0.76$ &  $0.72$ &  $0.36$ &  $0.39$ \\
%\hline
%interval$-$ &  $\rho$ & $-0.37 \pm 0.07$ &  $0.29 \pm 0.07$ &  $-0.25 \pm 0.09$ &  $0.25 \pm 0.10$ \\
%\hfil &  $p$ & $ 0.08$ &  $0.18$ &  $0.25$ &  $0.25$ \\
%\hline
%\end{tabular}
%\end{table}

\begin{table}[htb!]
\caption{Correlations with following cycle}
\label{tbl:corrforw}
\begin{tabular}{lcllll}     
\hline
\hfil & \hfil & amplitude$+$ & risetime$+$ & falltime$+$ & gap$+$ \\
\hline
amplitude & $\rho$ & $0.46 \pm 0.02$ &  $-0.21 \pm 0.06$ &  $0.39 \pm 0.05$ &  $-0.06 \pm 0.07$ \\
\hfil &  $p$ & $0.03$ &  $0.34$ &  $0.07$ &  $0.80$ \\
\hline
risetime & $\rho$ &  \hfil &  $0.05 \pm 0.10$ &  $-0.27 \pm 0.07$ &  $-0.06 \pm 0.13$ \\
\hfil &  $p$ & \hfil &  $0.83$ &  $0.21$ &  $0.37$ \\
\hline
falltime &  $\rho$ & \hfil & \hfil &  $0.20 \pm 0.07$ &  $0.19 \pm 0.07$ \\
\hfil &  $p$ & \hfil &  \hfil &  $0.36$ &  $0.39$ \\
\hline
gap &  $\rho$ & \hfil & \hfil  &  \hfil &  $0.25 \pm 0.10$ \\
\hfil &  $p$ & \hfil &  \hfil & \hfil &  $0.25$ \\
\hline
\end{tabular}
\end{table}

\begin{table}[htb!]
\caption{Correlations with previous cycle}
\label{tbl:corrback}
\begin{tabular}{lcllll}     
\hline
\hfil & \hfil & amplitude & risetime & falltime & gap \\
\hline
amplitude$-$ & $\rho$ & $0.46 \pm 0.02$ &  \hfil &  \hfil &  \hfil \\
\hfil &  $p$ & $0.03$ &  \hfil &  \hfil &  \hfil \\
\hline
risetime$-$ & $\rho$ &  $-0.17 \pm 0.08$ &  $0.05 \pm 0.10$ &  \hfil &  \hfil \\
\hfil &  $p$ & $ 0.44$ &  $0.83$ &  \hfil &  \hfil \\
\hline
falltime$-$ &  $\rho$ & $ 0.07 \pm 0.05$ &  $-0.08 \pm 0.06$ &  $0.20 \pm 0.07$ &  \hfil \\
\hfil &  $p$ & $ 0.76$ &  $0.72$ &  $0.36$ &  \hfil \\
\hline
gap$-$ &  $\rho$ & $-0.37 \pm 0.07$ &  $0.29 \pm 0.07$ &  $-0.25 \pm 0.09$ &  $0.25 \pm 0.10$ \\
\hfil &  $p$ & $ 0.08$ &  $0.18$ &  $0.25$ &  $0.25$ \\
\hline
\end{tabular}
\end{table}

We thus consider the relation between parameters in consecutive cycles so that they may be exploited to predict sunspot numbers of an entirely new cycle.  We have carried out numerous checks of the parameter combinations.  In Figure~\ref{fig:amprise} we display two such relations between the fitted values (posterior means) based on the first-stage model (Equation~(\ref{eq:stage1})): the amplitude of the next cycle [$c_{i+1}$] against the amplitude of the current cycle [$c_i$] (left plot), and against the time from maximum to the start of the next cycle [$t^{(i+1)}_{0}-t^{(i)}_{max}$] (right plot). 

\begin{figure}[htb!]
\caption{Relationships between the next amplitude $c_{i+1}$ and the current amplitude $c_i$, and between $c_{i+1}$ and 
$t_0^{(i+1)}-t_{\rm max}^{(i)}$.
Reported are posterior means based on the first-stage model in Equation~(\ref{eq:stage1}).
%Uncertainties in these quantities are represented by the vertical and horizontal bars ($\pm$ 1 posterior standard 
%deviation) centered at each data point.
}
\label{fig:amprise}
\begin{center} 
\psfrag{amp2}{\small $c_{i+1}$}
\psfrag{amp1}{\small $c_i$}
\psfrag{maxtostart}{\small $t_0^{(i+1)}-t_{\rm max}^{(i)}$}
  \includegraphics[width=2.7in, angle=270]{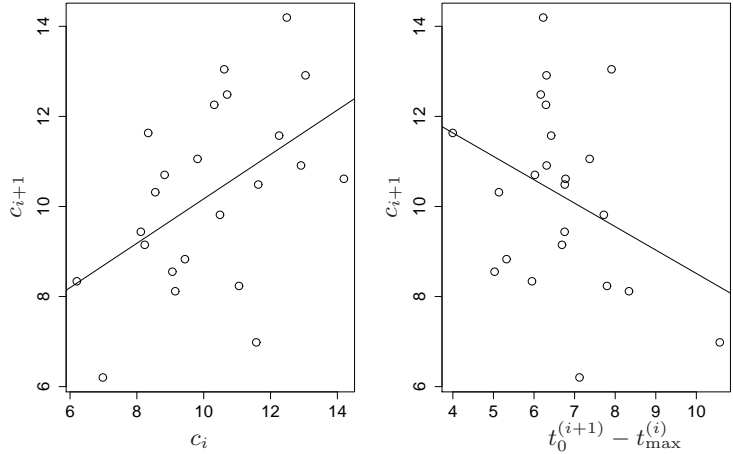}
\end{center}
\end{figure}

Based on the above exploratory analyses, we consider a combination of parameters to define a statistical model for predicting the next cycle parameters.
First, we enhance the predictive power of the positive correlation between successive amplitudes by combining it with the negative correlation shown on the right plot of Figure~\ref{fig:amprise},
\begin{equation}
c_{i+1}\sim \delta_1+\gamma_1\frac{c_i}{t_0^{(i+1)}-t_{\rm max}^{(i)}}+N(0, \sigma_1^2).
\label{eqn1}
\end{equation}

Next, we observe the following form of Waldmeier effect relating the rising time of each cycle to its amplitude, 
\begin{equation}
\label{eqn2}
t_{\rm max}^{(i+1)}-t_0^{(i+1)}\sim \delta_2+\gamma_2 c_{i+1}+N(0, \sigma_2^2).
\end{equation}
This relationship is illustrated in the middle panel in Figure~\ref{fig:cycrel}, again based on the first-stage model in Equation~(\ref{eq:stage1}). 

Finally, we incorporate the relatively weak correlation between the amplitude and the duration of the declining phase of each cycle, 
\begin{equation}
t_1^{(i+1)}-t_{\rm max}^{(i+1)}\sim \delta_3+\gamma_3 c_{i+1}+N(0, \sigma_3^2).
\label{eqn3}
\end{equation}
Figure~\ref{fig:cycrel} illustrates Equations~(\ref{eqn1}), (\ref{eqn2}) and (\ref{eqn3}).  
%Note the potential nonlinearity between $t_1^{(i+1)}-t_{\rm max}^{(i+1)}$ and $c_{i+1}$ from the third panel of Figure~\ref{fig:cycrel}.  
Although the third panel of Figure~\ref{fig:cycrel} exhibits a relatively weak relationship, we note that Equation~(\ref{eqn3}) is not needed for predicting the timing and amplitude of the next solar maximum, which is often the main goal. 

Note that only one of these relations explicitly connects the parameters of one cycle with that of the next (Equation (\ref{eqn1})).  The other two (Equations~(\ref{eqn2}) and (\ref{eqn3})) are based on parameter correlations within a cycle, but are computed for the following cycle.  This sequence, of first computing the amplitude of the following cycle, and using that to compute the rise and fall times of that cycle, is an important facet of the predictive capacity of our model.  The calculations cannot be carried out in a different order. 

\begin{figure}[htb!]
\caption{Three linear relationships useful for prediction, based on fitting the first-stage model in Equation~(\ref{eq:stage1}).
}
\label{fig:cycrel}
\begin{center}
\psfrag{amp2}{\small $c_{i+1}$}
\psfrag{amplitude/maxtostart}{\small $c_i/(t_0^{(i+1)}-t_{\rm max}^{(i)})$}
\psfrag{timetofall}{\small $t_1^{(i+1)}-t_{\rm max}^{(i+1)}$}
\psfrag{timetorise}{\small $t_{\rm max}^{(i+1)}-t_0^{(i+1)}$}
  \includegraphics[width=3in, angle=270]{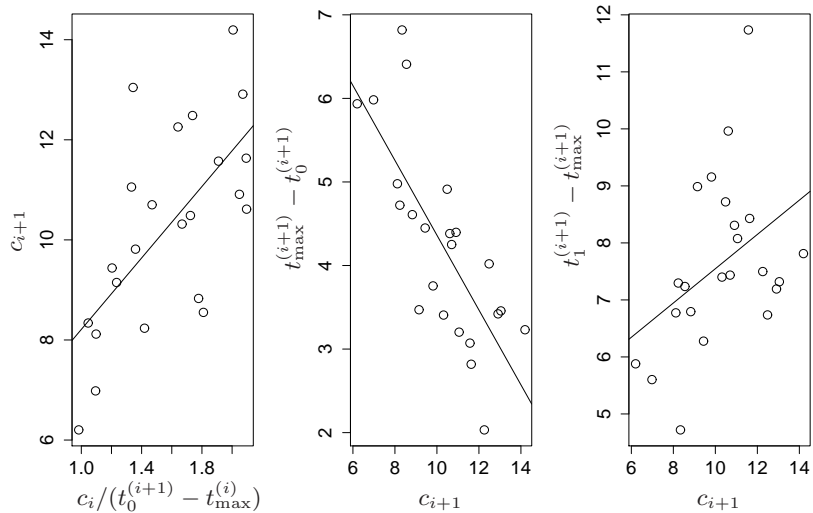}
\end{center}
\end{figure}

It should be emphasized that, because of the large amount of raw SSN observations, the quantities $c_i,\ t_{\rm max}^{(i)},\ t_1^{(i)},\ i=0,\ldots, 23,$ and $t_0^{(i)},\ i=1,\ldots, 24,$ are well constrained by the first stage model in Equation~(\ref{eq:stage1}).
Nonetheless, their fitted values (posterior means) do not account for the uncertainties and are used only for the purpose of illustration.
The mathematical forms of the relationships in Equations~(\ref{eqn1})\,--\,(\ref{eqn3}) are found by examining the posterior means of these parameters, but ultimately the parameters will be fit using the raw data.
In Section~\ref{sec:fitting} we describe a two-level joint modeling approach which automatically accounts for uncertainty in these quantities. 

We have also explored possible correlations between separated cycles (i.e., between the $k^{\rm th}$ and the $(k{\pm}2)^{\rm th}$ cycles).
However, given the characteristics of Cycle $k-1$, we find no evidence for a conclusive dependence of Cycle $k$ on Cycle $k{\pm}2$.
Such dependences are weak and have little predictive value.
Therefore we focus on lag-one dependence.

\subsection{Model Fitting}
\label{sec:fitting}

We adopt a Bayesian approach based on the relations found in the previous stage and carry out principled parameter estimations for the correlation coefficients.
We employ Markov chain Monte Carlo methods that fit both stages of the statistical model simultaneously.  This approach makes it easy to correctly account for the uncertainty both in estimating the parameters of the individual cycles and in modeling the relations among cycles. 
%For parameter estimation we adopt a Bayesian approach and employ Markov chain Monte Carlo methods that fit both stages of the statistical model simultaneously.
%The resulting simulations can also be used to predict future cycle characteristics, yielding an estimated maximum smoothed monthly SSN of 78 $\pm$ 18, to occur in April/May 2014 $\pm$ 8 months, for Cycle 24, using one sigma ($68\%$) error bars. 

\subsubsection{Bayesian Hierarchical Models}
The results in Section~\ref{sec:statmodel} are obtained by fitting Equation~(\ref{eq:stage1}) using a Bayesian approach.
In this section we describe the model fitting procedure as well as how the Bayesian approach can be used to fit both stages (Equations~(\ref{eq:stage1}) and (\ref{eqn1})\,--\,(\ref{eqn3})) together as a single more coherent statistical model.
See Gelman {\it et al.} (2004) for an introduction to Bayesian hierarchical modeling and the associated computational strategies.
See van Dyk {\it et al.} (2001) for the use of Bayesian methods in the context of highly structured models for spectral analysis in high energy astrophysics, and Esch {\it et al.} (2004) in the context of multiscale image reconstruction.
In a Bayesian analysis, the likelihood function [$p(Y|\theta)$] is combined with a prior distribution [$p(\theta)$] to form a posterior distribution,
$$p(\theta|Y)\propto p(Y|\theta) p(\theta),$$
and all inference is derived from this posterior distribution.
One may build further structures on the prior by introducing some hyper-parameter [$\eta$] with its own prior [$p(\eta)$] (the hyper-prior), and replacing $p(\theta)$ with $p(\theta|\eta)$.
This leads to a two-level model.
By Bayes' Theorem, the joint distribution of $(\theta, \eta)$ given the data $Y$ can be written as
$$p(\theta, \eta|Y)\propto p(Y|\theta, \eta) p(\theta|\eta) p(\eta),$$
where $p(Y|\theta, \eta)$ is the likelihood of observed data (the first stage), $p(\theta|\eta)$ is the second stage distribution for $\theta$, and $p(\eta)$ is the hyper-prior.
Inference concerning $\theta$, for example, is based on its marginal posterior distribution $p(\theta|Y)= \int p(\theta, \eta|Y)\, {\rm d}\eta$. 

In our case $Y=\{Y_t\}$ and $\theta$ is the collection of parameters 
$$\theta=(T_0, T_{\rm max}, T_1, C, \alpha, \beta, \sigma^2).$$
The likelihood function $p(Y|\theta)$ is determined by Equation~(\ref{eq:stage1}).
We impose independent uniform prior distributions on $\beta$ and $\log\sigma$, as is commonly done for such parameters in a first-stage regression model.
For each of $\alpha_1, \alpha_2$, a uniform prior on $[1,3]$ is used to allow for a flexible range of cycle shapes.
When fitting the first stage model (\ref{eq:stage1}) alone, as in Section~\ref{sec:stage1}, noninformative uniform priors are assigned to the other components of $\theta$, {\it i.e.} $T_0, T_{\rm max}, T_1$, and $C$, subject to natural constraints on their ranges.
In this section we fit the two stages jointly, and these components of $\theta$ are linked together by Equations~(\ref{eqn1})\,--\,(\ref{eqn3}).
In addition, we express the starting point of the next cycle $t_0^{(i+1)}$, given the end point of the current cycle $t_1^{(i)}$, as 
\begin{equation}
\label{t0t1}
t_0^{(i+1)}\sim t_1^{(i)} + N(0, \tau^2).
\end{equation}
The parameter $\tau^2$ regulates how far apart $t_0^{(i+1)}$ and $t_1^{(i)}$ are allowed to be.
Allowing $t_0^{(i+1)}$ to be different from $t_1^{(i)}$ offers additional flexibility.
In summary, Equations~(\ref{eqn1})\,--\,(\ref{t0t1}) specify the distribution of $T_0, T_{\rm max}, T_1$, and $C$ given the hyper-parameters $\eta = (\tau^2, \gamma_j, \delta_j, \sigma^2_j,\ j=1,2,3).$
For these hyper-parameters we use independent non-informative priors (specifically, uniform priors) on $\tau, \gamma_j, \delta_j, \sigma_j$, $j=1,2,3$. 

\subsubsection{Model Fitting with MCMC}
Markov chain Monte Carlo (MCMC) techniques are used to draw samples from the posterior distribution, which can then be summarized as point estimates and error bars of the parameters of interest ({\it e.g.} van Dyk {\it et al.}, 2001 and Park {\it et al.}, 2008).
In general, if we can simulate random samples $\theta^{(1)},\ldots, \theta^{(L)}$ from $p(\theta|Y)$, the target posterior distribution, then inferences for quantities of interest can be derived by examining the empirical distribution of these samples.
For example, suppose $\theta$ is one-dimensional and has a symmetric and unimodal posterior distribution, then a natural point estimate of $\theta$ is the posterior mean, which can be approximated by $L^{-1}\sum_{l=1}^L \theta^{(l)}$, the empirical average of the posterior sample.
The posterior standard deviation serves as a natural one-sigma error bar, and can be approximated by the sample standard deviation of $\theta^{(1)},\ldots, \theta^{(L)}$. 

In high-dimensional situations when direct simulation from the posterior distribution is difficult, as is the case of our analysis, one may adopt an MCMC approach, which constructs a Markov chain with the desired posterior distribution $p(\theta|Y)$ as its stationary distribution.
After the Markov chain reaches equilibrium, the iterations of $\theta$ can then be used as (dependent) samples from the target distribution.
Two well-known methods for constructing such Markov chains are the Metropolis--Hastings (M--H) algorithm (Metropolis {\it et al.}, 1953; Hastings, 1970) and the Gibbs sampler (Geman and Geman, 1984; Gelfand and Smith, 1990).
In an M--H strategy, to obtain the next iteration of the Markov chain, a sample is drawn from a proposal distribution, and it is accepted or rejected according to a certain probability so that the target distribution is preserved.
In Gibbs sampling, the parameter $\theta$ is partitioned into several components, and at each iteration, we update each component in turn by drawing from its conditional posterior distribution given all other components.
The actual MCMC algorithm used for fitting our two-stage model is a hybrid algorithm that cycles through the coordinates of the parameter vector in a Gibbs sampling fashion but uses an M--H strategy for each conditional draw.
The algorithm is carefully monitored; several Markov chains from different starting values are run to ensure that they reach the same target distribution. 

\end{section}

\begin{section}{Joint Fitting: Results and Discussion}
\label{sec:discuss}

We now discuss the results from the fitted hierarchical model. This is a more coherent analysis than the separate fitting of the two stages as described in Section~\ref{sec:statmodel}.  Thus, the results in this section represent our final estimates. 

\subsection{Cycle-to-Cycle Dependencies}
\label{sec:cyccor}

Of particular interest are the estimates of the three second-stage relationships.
These estimates are summarized in Table~\ref{tbl:result}, based on an MCMC calculation that fits the two stages jointly, using available data from January 1749 to October 2011.
As noted before, because of the large number of observations, cycle characteristics such as $c_i, t_0^{(i)}, t_{\rm max}^{(i)}$, and $t_1^{(i)}$ are well constrained by the first-stage model (Equation~(\ref{eq:stage1})), and hence the estimates reported in Table~\ref{tbl:result} are quite close to those based on an ordinary least squares (OLS) regression using fixed $c_i, t_0^{(i)}, t_{\rm max}^{(i)}, t_1^{(i)}$.
For example, with $c_i, t_0^{(i)}, t_{\rm max}^{(i)}$ fixed at their respective posterior means, the OLS estimates (standard errors) of $\delta_2$ and $\gamma_2$ are $8.5(\pm 0.8)$ and $-0.43(\pm 0.08)$, respectively.
The standard errors in Table~\ref{tbl:result} are slightly larger because they account for the extra uncertainty in estimating $c_i, t_0^{(i)}, t_{\rm max}^{(i)}$, and $t_1^{(i)}$.
The quality of the predictive relationships is shown in Figure~\ref{fig:ratio_predobs}, where we display the ratios of the values predicted from the previous cycle to those measured for that cycle, together with 1-$\sigma$ prediction error bars.  Here the posterior mean estimates (as in Table~\ref{tbl:fitparam}) are regarded as measured or estimated values; the predicted ones are computed using the equations in Table~\ref{tbl:result}.  The prediction error bars are computed by simulation.  Note that the coefficients $\hat{\delta}_i,\ \hat{\gamma}_i,\ i=1,2,3,$ are estimated using all cycles (including those that come after the one being predicted). 

\begin{figure}[htb!]
\caption{Ratios of predicted parameter values to measured values.  The values predicted for the parameters in each
cycle (amplitude, rise time and fall time) based on the parameters of the previous cycle, are compared with the values
directly estimated for that cycle.}
\label{fig:ratio_predobs}
\begin{center}
\includegraphics[width=2.4in, angle=270]{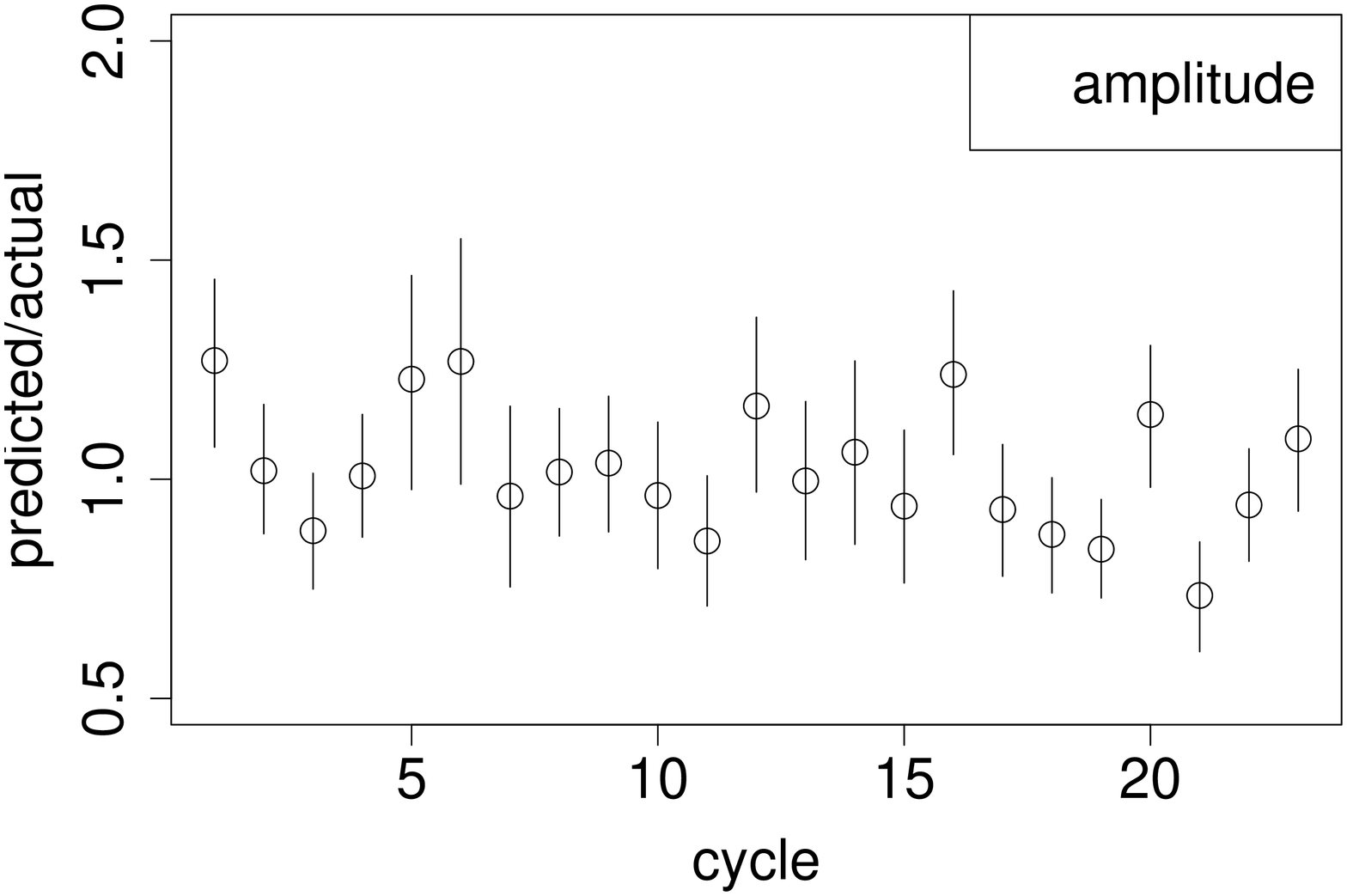}
\includegraphics[width=2.4in, angle=270]{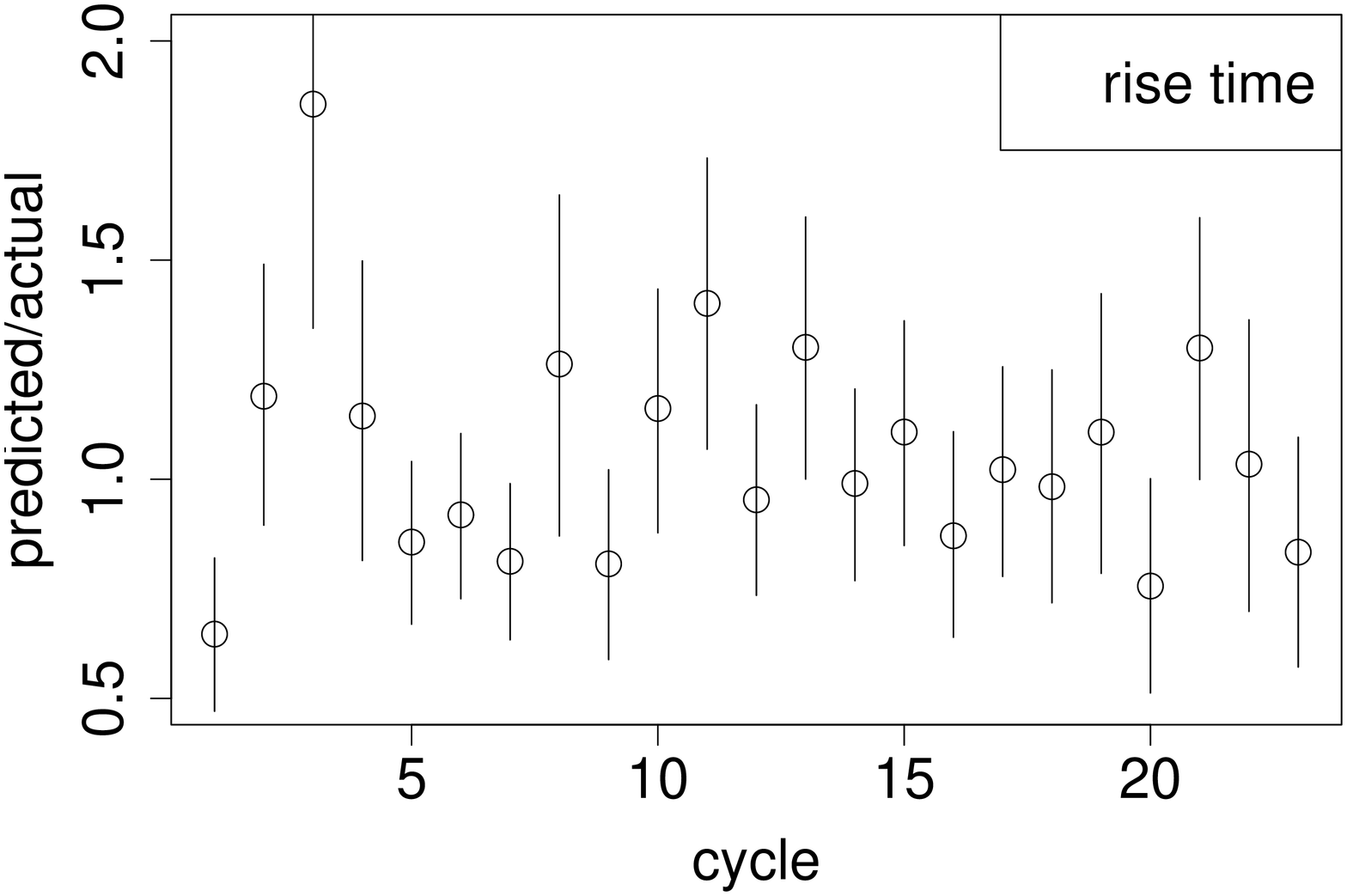}
\includegraphics[width=2.4in, angle=270]{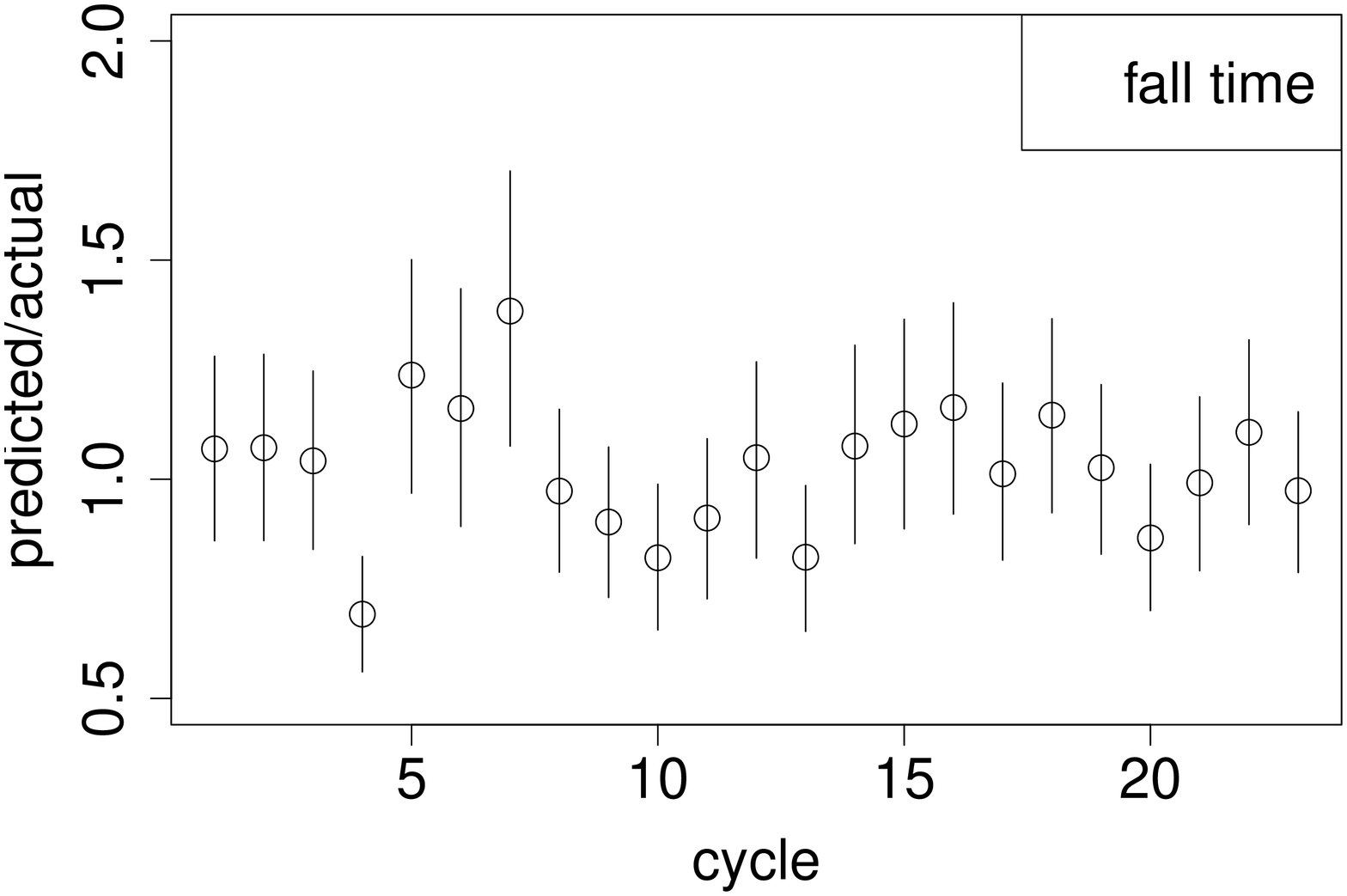}
\end{center}
\end{figure}

\begin{table}[htb!]
\caption{Summary of three relationships and their parameter estimates when fitting the two stages jointly.
The fitted values are posterior means and the standard errors are posterior standard deviations.
}
\label{tbl:result}
\begin{center}
\begin{tabular}{lll}
\hline
Cycle parameter & Relationship                                                        & Fitted value (Std.\ err.) \\
\hline
Amplitude       & $c_{i+1}\sim \delta_1+\gamma_1 c_i/(t_0^{(i+1)}-t_{\rm max}^{(i)})$ &  $\hat{\delta}_1=4.1\, (\pm 1.5)$           \\
                &                                                                     &  $\hat{\gamma}_1=3.9\, (\pm 1.0)$           \\
\hline
Time to maximum & $t_{\rm max}^{(i+1)}-t_0^{(i+1)}\sim \delta_2+\gamma_2 c_{i+1}$     &  $\hat{\delta}_2=8.5\, (\pm 1.0)$           \\
                &                                                                     &  $\hat{\gamma}_2=-0.43\, (\pm 0.09)$          \\
\hline
Time to minimum & $t_1^{(i+1)}-t_{\rm max}^{(i+1)}\sim \delta_3+\gamma_3 c_{i+1}$     &  $\hat{\delta}_3= 4.3\, (\pm 1.5)$           \\
                &                                                                     &  $\hat{\gamma}_3=0.31\, (\pm 0.15)$          \\
\hline
\end{tabular}
\end{center}
\end{table}

\subsection{Predictions for Cycle 24}
\label{sec:cyc24}

One advantage of a Bayesian hierarchical model in this context is that, once we obtain the samples from the posterior distribution, prediction of the characteristics of the current (incomplete) cycle is obtained automatically.
Table~\ref{tbl:cyc24} reports summaries of the posterior inference for Cycle 24, using data up to November 2008, May 2010, and October 2011, respectively.  Based on data up to October 2011, Cycle 24 is estimated to rise to maximum in January -- February 2014 $\pm$ six months, with a maximum smoothed monthly sunspot number of $97\pm 15$, where the estimates are posterior means and the error bars are posterior standard deviations.  (The maximum smoothed SSN, or the expected SSN at solar maximum, is $(\beta+c_i)^2+\sigma^2$, after accounting for the square-root transformation in Equation~(\ref{eq:stage1}).)  It is likely to be a weak cycle with a longer-than-usual (an expected 12.1 years) total duration, although the uncertainty associated with this estimate is fairly large.  We observe that the estimated maximum smoothed sunspot number is relatively stable across the three analyses.  The large error bars associated with the November 2008 analysis highlight the inherent difficulty in making predictions before or at the onset of a cycle. 

\begin{table}[htb!]
\caption{Cycle 24 predictions based on MCMC fitting of the two stages jointly.
Fitted values are posterior means and standard errors are posterior standard deviations.
Max.\ SSN refers to the smoothed monthly average sunspot number at the peak of Cycle 24.
Time to rise [years] is defined as $t_{\rm max}^{(24)}- t_0^{(24)}$.
}
\label{tbl:cyc24}
\begin{center}
\begin{tabular}{rrrrrr}
\hline
          & $c_{24}$           & Max.\ SSN       & Time of max.\ [yrs]             & Time to rise          &  Cycle length        \\    

Nov\,08   & $9.0\pm 1.6$       & $96\pm 32$      &  Mar $2013 \pm 0.98$    &  $4.7\pm 1.1$      &  $11.9\pm 1.7$     \\ 

May\,10   & $8.2\pm 1.2$       & $77\pm 21$     &  May $2014\pm 0.69$        &  $5.3\pm 0.84$      &  $12.1\pm 1.6$     \\ 

Oct\,11   & $9.3\pm 0.77$      & $97\pm 15$     &  Jan/Feb $2014\pm 0.48$    &  $4.8\pm 0.55$      &  $12.1\pm 1.5$     \\ 
\hline
\end{tabular}
\end{center}
\end{table}

Figure~\ref{fig:cyc24} illustrates the estimates of the averages of $Y_t$ in Equation~(\ref{eq:stage1}) (specifically, $(\beta + U_t)^2 + \sigma^2$). 
The solid curve represents the posterior mean while the upper (lower) dashed curve represents the $95\%$ ($5\%$) posterior quantile.
Note that estimates for time points in the past are well constrained because of the available data, but future predictions are much more variable.
The two-stage model is well suited for combining two pieces of information that have potential predictive power: sunspot number observations that clearly belong to Cycle 24, and the prescription of the second-stage model (Equations~(\ref{eqn1})\,--\,(\ref{eqn3})) which relates the characteristics of Cycle 24 to those of previous cycles.
Given the relatively few observations at the beginning of Cycle 24, the predictions are heavily influenced by the second-stage model.
As Cycle 24 progresses, direct observations will play a heavier role and the uncertainties 
associated with the predictions will diminish.
This is illustrated by the reduction in the uncertainty band in the bottom panel which includes 35 more months of observations (up to October 2011) compared to that in the top panel (up to November 2008).  (This reduction in uncertainty is also apparent from Table~\ref{tbl:cyc24}.)
When the more recent data are included, the predictions are more driven by direct observations from Cycle 24; the fitted values are similar, but the 90\% predictive intervals are appreciably narrower.
This shows that the latest data are reasonably consistent with the second-stage relationships, and combining the two stages shrinks the error bars. 
 
\begin{figure}[ht]
\caption{Predictions of Cycle 24 obtained by fitting the two stages jointly, using data up to November 2008 (top), May 2010 (middle) and October 2011 (bottom).  The posterior mean of monthly average sunspot numbers is shown as the solid curve, and the 5\% and 95\% posterior quantiles are shown as dashed curves.  The top figure illustrates predictions for a completely new cycle, and the bottom for one that is well in progress.  May 2010 is chosen as it lies half way in between November 2008 and October 2011.  Note that the uncertainty in the predictions is reduced considerably when more data from the current cycle are included. 
}
\label{fig:cyc24}
\begin{center}
  \includegraphics[width=2.4in, angle=270]{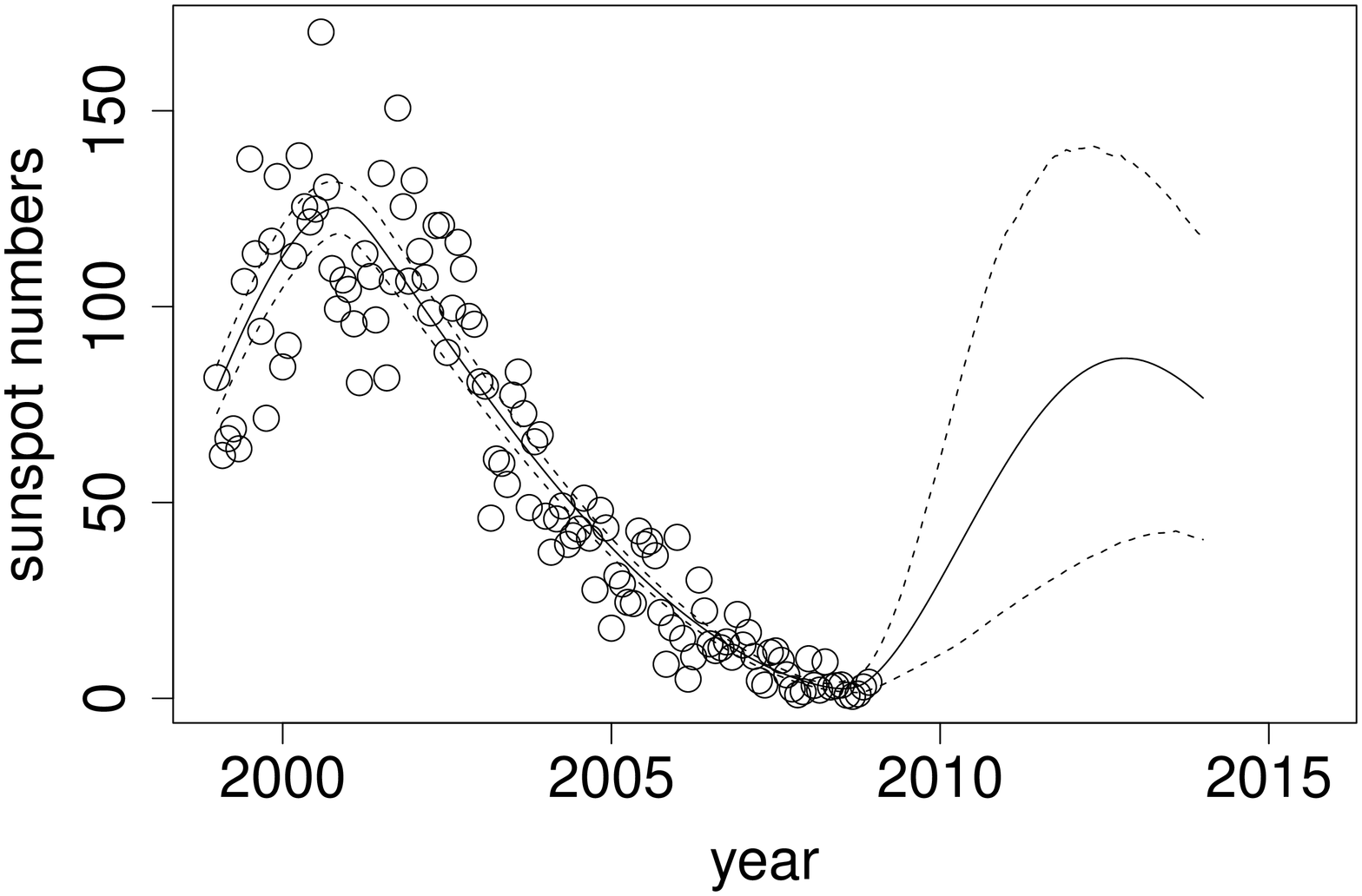}\\
  \includegraphics[width=2.4in, angle=270]{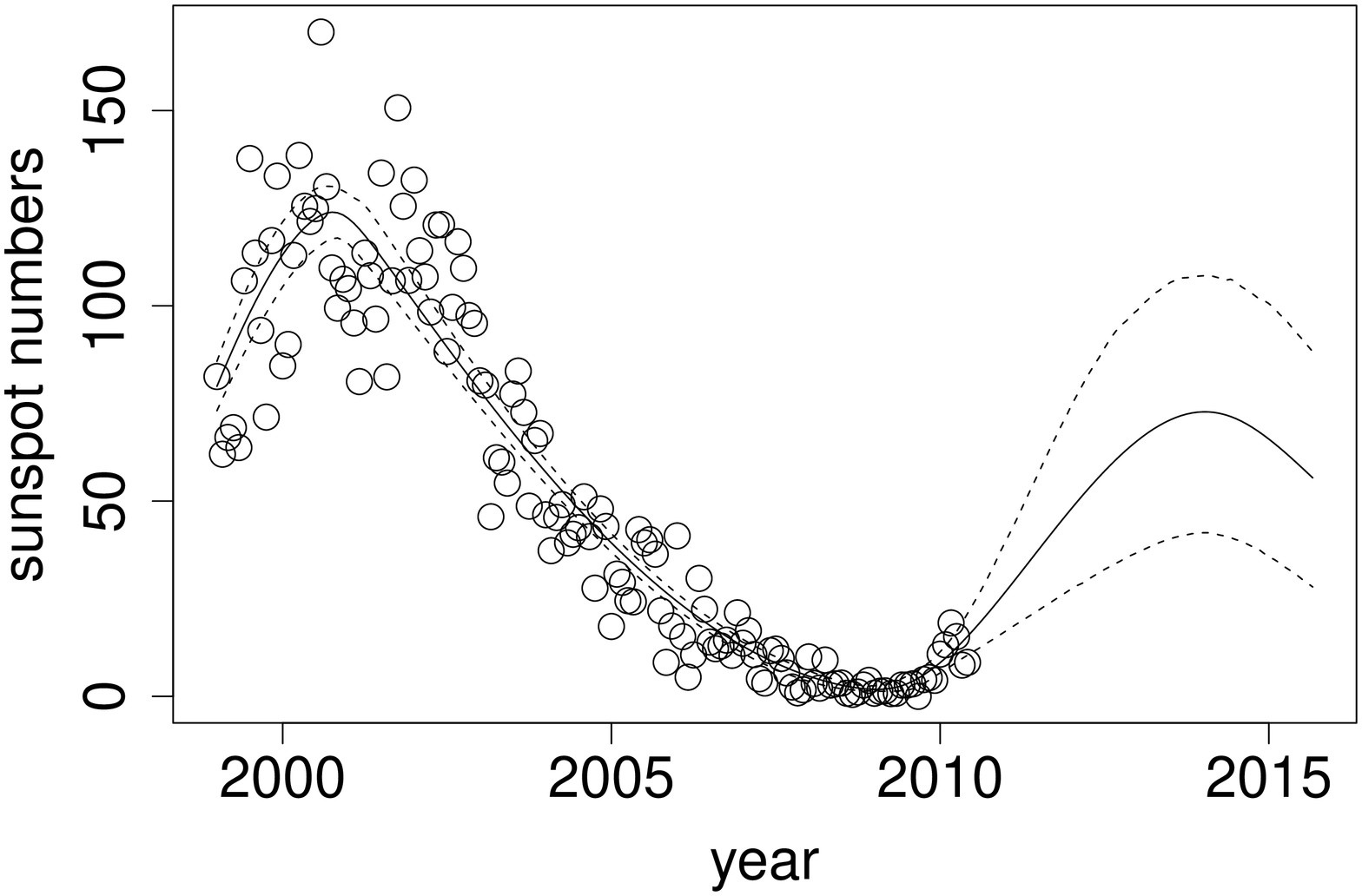}\\
  \includegraphics[width=2.4in, angle=270]{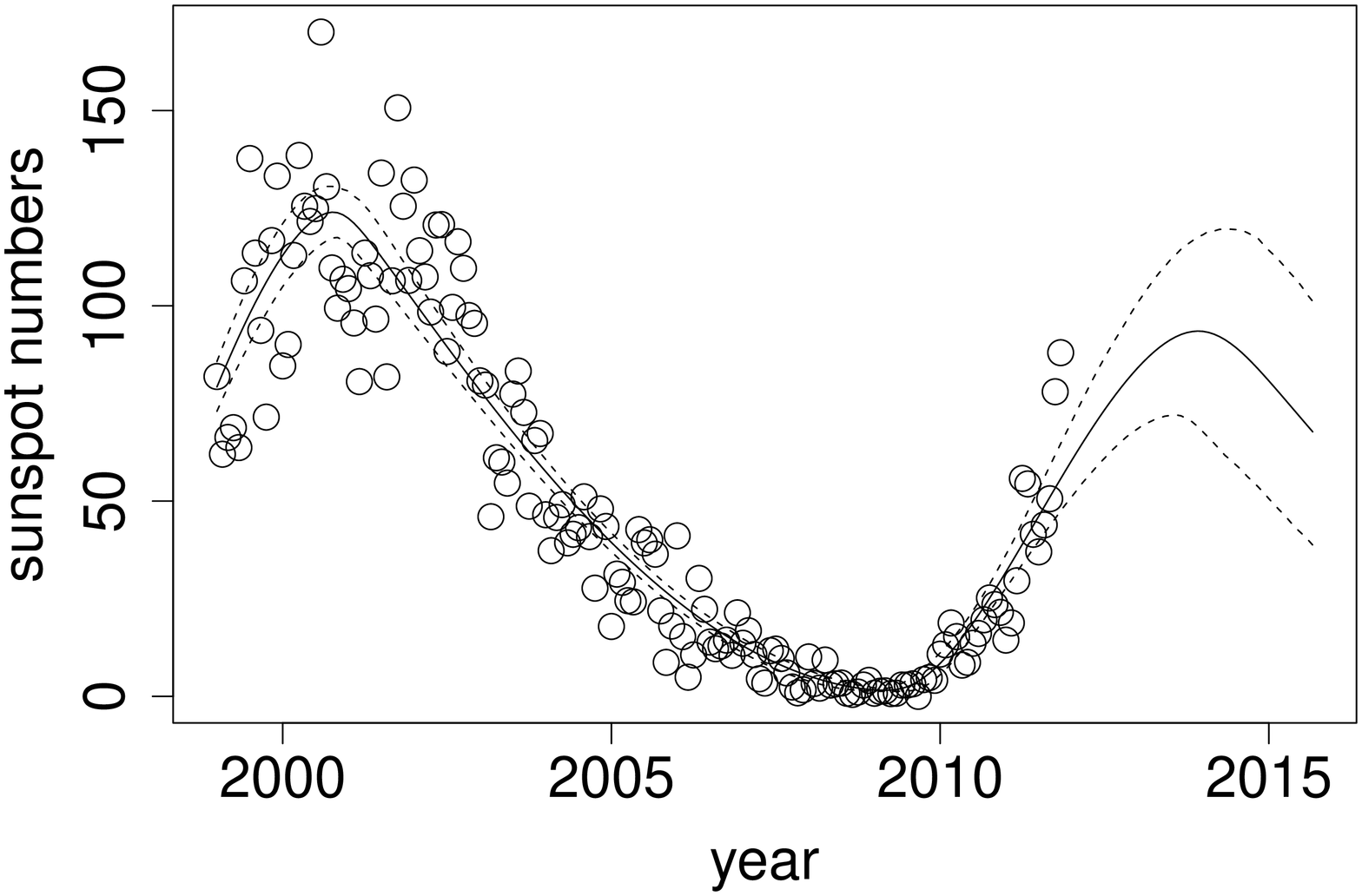}
\end{center}
\end{figure}

\end{section}

\begin{section}{Summary}
\label{sec:summary}

We have carried out a comprehensive statistical analysis of the sunspot record.
After suitably transforming the data to stabilize variance, we parameterized the shape of a cycle by its amplitude (maximum in the sunspot number), time to rise to maximum, time to fall to minimum, and the gap between its end and the start of the next cycle. 
By computing correlations between these parameters both within each cycle and between adjacent cycles, we have derived a set of three predictive relations.  These relations are ordered, {\it i.e.} sequential: amplitude must be predicted first before duration and rise time.
Correlations that depend on computing amplitude second are not robust, as they are subject to two influential points from early in the sunspot record (see also Vaquero and Trigo 2008).  Analyses carried out in a different order will thus lead to spurious results. 

These relations can be used to predict the values of the parameters for the following cycle.
We find that the best estimate for the peak in Cycle 24 is in early 2014, with an uncertainty of half a year.
The maximum in the smoothed sunspot number record is expected to be $\approx{97}\pm{15}$, and the cycle is expected to last $12.1{\pm}1.5$ years from the latest solar minimum (approximately November 2008).
These are in the middle of the range of predictions in the literature prior to the onset of Cycle 24, and are consistent with the current estimates of the cycle parameters. 

We have searched for, but do not find, any evidence for persistence beyond one cycle.
There is no predictive power beyond the cycle that follows; no correlations are present, and the cycles do not retain any memory.

We also find that the cycles do not ever vanish completely.
We find statistical evidence that the next cycle usually begins before the current cycle ends, as the gap between the end of a cycle and the start of a new one is usually negative.  Furthermore, we find that sunspots do not vanish entirely even if the
gap were positive; however, the data are not sufficient to tell whether this holds true even in the {\sl absence} of activity cycles.

\acknowledgements This work was supported by CXC NASA contract NAS 8-39073 (VLK) and NSF grants DMS 04-06085 and DMS 09-07522 (DvD, YY).

\end{section}

%-------------------------------------------------

%%%%%%%%%%%%%%%%%%%%%%%%%%%%%%%%%%%%%%%%%%%%%%%%%%%
%% Sections
%
% \section{}%\label{s:?} 

%% Figure 
%
% \begin{figure} 
% \centerline{\includegraphics[width=0.5\textwidth,clip=]{<fig.eps>}}
% \caption{}%\label{fig:?}
% \end{figure}

%% Table
%
% \begin{table}
% \caption{}%\label{tbl:?}
% \begin{tabular}{}     
% \hline
% \multicolumn{2}{c}{<>}
% <data>
% \hline
% \end{tabular}
% \end{table}

%%%%%%%%%%%%%%%%%%%%%%%%%%%%%%%%%%%%%%%%%%%%%%%%%%%%%%%%%%%%%%%%%%%%%%%%%%%
%% Appendix
%
% \appendix   

%%%%%%%%%%%%%%%%%%%%%%%%%%%%%%%%%%%%%%%%%%%%%%%%%%%%%%%%%%%%%%%%%%%%%%%%%%%
%% Acknowledgements
%
% \begin{acks}
%
% \end{acks}

%%% %%%%%%%%%%%%%%%%%%%%%%%%%%%%%%%%%%%%%%%%%%%%%%%%%%%%%%%%%%%
%% Bibliography
%
% Using BibTeX
%
% \bibliographystyle{spr-mp-sola}
% %\bibliographystyle{spr-mp-sola-cnd} %% Alternative style: no title, no concluding page
% \bibliography{<bib file>}  

\begin{thebibliography}{99}
%\bibitem{ALM}
%Aguirre, L.A., Letellier, C., \& Maquet, J., 2008, {\it Solar Phys.}, {\bf 249}, 103. 

%\bibitem{BSKF}
%Balmaceda, L.A., Solanki, S.K., Krivova, N.A., \& Foster, S., 2009, {\it JGRA}, {\bf 114}(A7), A07104. (arxiv: 0906.0942)

\bibitem{B05}
Benestad, R.E., 2005, {\it Geophys. Res. Lett.}, {\bf 32}, L15714.

%\bibitem{BS}
%Brandenburg, A., \& Spiegel, E., 2008, {\it AN}, submitted (arXiv:0801.2156)

\bibitem{BPS}
Bonev, B.P., Penev, K.M., Sello, S., 2003, {\it Astrophys. J.}, {\bf 605}, L81.

\bibitem{C}
Charbonneau, P., 2007, {\it Adv. Sp. Res.}, {\bf 39}(11), 1661.

\bibitem{CD}
Charbonneau, P., Dikpati, M., 2000, {\it Astrophys. J.}, {\bf 543}, 1027. 

\bibitem{Cho}
Choudhuri, A.R., 1992, {\it Astron. Astrophys}, {\bf 253}, 277.

\bibitem{CCJ}
Choudhuri, A.R., Chatterjee, P., Jiang, J. 2007, {\it Phys. Rev. Lett.}, {\bf 98}, 131103.

\bibitem{DTG}
Dikpati, M., de Toma, G., Gilman, P.A., 2006, {\it Geophys. Res. Lett.}, {\bf 33}, L05102.

\bibitem{DG}
Dikpati, M., Gilman, P.A., 2006, {\it Astrophys. J.}, {\bf 649}, 498.

\bibitem{EMC2}
Esch, D.N., Connors, A., Karovska, M., van Dyk, D.A., 2004, {\it Astrophys. J.}, {\bf 610}, 1213.

\bibitem{GS}
Gelfand, A.E., Smith, A.F.M., 1990, {\it J. Amer. Statist. Assoc.} {\bf 85}, 398--409.

\bibitem{GCSR}
Gelman, A., Carlin, J.B., Stern, H.S., Rubin, D.B., 2004, {\it Bayesian Data Analysis}, 2nd ed.  London: CRC Press.

\bibitem{GG}
Geman, S., Geman, D., 1984, {\it IEEE Trans. Pattern Analysis Machine Intelligence} {\bf 6}, 721--741. 

\bibitem{Gil}
Gil-Alana, L.A., 2009, {\it Solar Phys.}, {\bf 257}, 371. \\
ADS: \href{http://adsabs.harvard.edu/abs/2009SoPh..257..371G}{\textsf{2009SoPh..257..371G}},
DOI: \href{http://dx.doi.org/10.1007/s11207-009-9390-1}{\textsf{10.1007/s11207-009-9390-1}}.

%\bibitem{Gre}
%Greenkorn, R.A., 2009, {\it Solar Phys.}, {\bf 255}, 301. 

\bibitem{Hast}
Hastings, W.K., 1970, {\it Biometrika} {\bf 57}, 97--109. 

\bibitem{HW}
Hathaway, D., Wilson, R.M., 2006, {\it Geophys. Res. Lett.}, {\bf 33}, L18101. 

\bibitem{HWR94}
Hathaway, D., Wilson, R.M., Reichmann, D. J., 1994, {\it Solar Phys.} {\bf 151}, 177. \\
ADS: \href{http://adsabs.harvard.edu/abs/1994SoPh..151..177H}{\textsf{1994SoPh..151..177H}},
DOI: \href{http://dx.doi.org/10.1007/BF00654090}{\textsf{10.1007/BF00654090}}. 

\bibitem{HWR02}
Hathaway, D., Wilson, R.M., Reichmann, E.J., 2002, {\it Solar Phys.} {\bf 211}, 357. \\
ADS: \href{http://adsabs.harvard.edu/abs/2002SoPh..211..357H}{\textsf{2002SoPh..211..357H}},
DOI: \href{http://dx.doi.org/10.1023/A:1022425402664}{\textsf{10.1023/A:1022425402664}}.

\bibitem{H2010}
Hill, F., Howe, R., Komm, R., Hern\'andez, I.G., Kholikov, S., Leibacher, J., 2010, {\it Astrophysical Dynamics: From Stars to Galaxies}, Proc.\ IAU Symp. ({\it Editors}: Brummell, N.H., Brun, A.S., Miesch, M.S., Ponty, Y.), {\bf 271}, 15, Cambridge University Press, Cambridge, UK.

\bibitem{Hud}
Hudson, H., 2007, {\it Astrophys. J.}, {\bf 663}, L45. 

\bibitem{K11}
Kakad, B., 2011, {\it Solar Phys.}, {\bf 270}, 393. \\
ADS: \href{http://adsabs.harvard.edu/abs/2011SoPh..270..393K}{\textsf{2011SoPh..270..393K}},
DOI: \href{http://dx.doi.org/10.1007/s11207-011-9726-5}{\textsf{10.1007/s11207-011-9726-5}}. 

\bibitem{K01}
Kane, R.P., 2001, {\it Solar Phys.}, {\bf 202}, 395. \\
ADS: \href{http://adsabs.harvard.edu/abs/2001SoPh..202..395K}{\textsf{2001SoPh..202..395K}},
DOI: \href{http://dx.doi.org/10.1023/A:1012211803591}{\textsf{10.1023/A:1012211803591}}.

\bibitem{K08}
Kane, R.P., 2008, {\it Solar Phys.}, {\bf 248}, 203. \\
ADS: \href{http://adsabs.harvard.edu/abs/2008SoPh..248..203K}{\textsf{2008SoPh..248..203K}},
DOI: \href{http://dx.doi.org/10.1007/s11207-008-9125-8}{\textsf{10.1007/s11207-008-9125-8}}.

%\bibitem{K08}
%Kane, R.P., 2008, {\it Journal of Atmospheric and Solar-Terrestrial Physics}, {\bf 70}, 1533. 

\bibitem{Ken}
Kendall, M., 1975, {\it Rank Correlation Methods}, Griffin, London, UK. 

\bibitem{Metro}
Metropolis, N., Rosenbluth, A.W., Rosenbluth, M.N., Teller, A.H., Teller, E., 1953, {\it J. Chem. Phys.} {\bf 21}, 1087--1092.

\bibitem{NW12} 
Noble, P.L., Wheatland, M.S., 2012, {\it Solar Phys.}, {\bf 276}, 363. \\
ADS: \href{http://adsabs.harvard.edu/abs/2012SoPh..276..363N}{\textsf{2012SoPh..276..363N}},
DOI: \href{http://dx.doi.org/10.1007/s11207-011-9884-5}{\textsf{10.1007/s11207-011-9884-5}}.

\bibitem{PvDS}
Park, T., van Dyk, D.A., Siemiginowska, A., 2008, {\it Astrophys. J.}, {\bf 688}, 807. 

\bibitem{P}
Pesnell, W.D., 2008, {\it Solar Phys.}, {\bf 252}, 209. \\
ADS: \href{http://adsabs.harvard.edu/abs/2008SoPh..252..209P}{\textsf{2008SoPh..252..209P}},
DOI: \href{http://dx.doi.org/10.1007/s11207-008-9252-2}{\textsf{10.1007/s11207-008-9252-2}}.

\bibitem{Ramesh}
Ramesh, K.B., Lakshmi, N.B., 2012, {\it Solar Phys.}, {\bf 276}, 395. \\
ADS: \href{http://adsabs.harvard.edu/abs/2012SoPh..276..395R}{\textsf{2012SoPh..276..395R}}, 
DOI: \href{http://dx.doi.org/10.1007/s11207-011-9866-7}{\textsf{10.1007/s11207-011-9866-7}}.

%\bibitem{R}
%Ramos, A.A., 2007, {\it A\&A}, {\bf 472}, 293. 

\bibitem{SA}
Sabarinath, A., Anilkumar, A.K., 2008, {\it Solar Phys.}, {\bf 250}, 183. \\ 
ADS: \href{http://adsabs.harvard.edu/abs/2008SoPh..250..183S}{\textsf{2008SoPh..250..183S}},
DOI: \href{http://dx.doi.org/10.1007/s11207-008-9209-5}{\textsf{10.1007/s11207-008-9209-5}}.

\bibitem{Sch}
Sch\"{u}ssler, M., 2007, {\it Astron. Nachr.}, {\bf 328}, 1087. 

\bibitem{SUKSB}
Solanki, S.K., Usoskin, I.G., Kromer, B., Sch\"{u}ssler, M., Beer, J., 2004, {\it Nature}, {\bf 431}, 1084.

\bibitem{S10}
Svalgaard, L., 2010, arXiv:1008.4832

\bibitem{USK}
Usoskin, I.G., Solanki, S.K., Kovaltsov, G.A., 2007, {\it Astron. Astrophys.}, {\bf 471}, 301. 

\bibitem{vanDyk} 
Van Dyk, D.A., Connors, A., Kashyap, V.L., Siemiginowska, A., 2001, {\it Astrophys. J.}, {\bf 548}, 224. 

\bibitem{Vaq}
Vaquero, J.M., Trigo, R.M., 2008, {\it Solar Phys.}, {\bf 250}, 199.\\ 
ADS: \href{http://adsabs.harvard.edu/abs/2008SoPh..250..199V}{\textsf{2008SoPh..250..199V}}, 
DOI: \href{http://dx.doi.org/10.1007/s11207-008-9211-y}{\textsf{10.1007/s11207-008-9211-y}}.

\bibitem{Vo}
Volobuev, D.M., 2009, {\it Solar Phys.}, {\bf 258}, 319.\\ 
ADS: \href{http://adsabs.harvard.edu/abs/2009SoPh..258..319V}{\textsf{2009SoPh..258..319V}}, 
DOI: \href{http://dx.doi.org/10.1007/s11207-009-9429-3}{\textsf{10.1007/s11207-009-9429-3}}.

\bibitem{W35a}
Waldmeier, M., 1935, {\it Astron. Mitt. Eidgen. Sternw. Z\"{u}rich}, {\bf 14}, 105. 

\bibitem{W35b}
Waldmeier, M., 1939, {\it Astron. Mitt. Eidgen. Sternw. Z\"{u}rich}, {\bf 14}, 470. 

\bibitem{W71}
Waldmeier, M., 1971, {\it Astron. Mitt. Eidgen. Sternw. Z\"{u}rich}, {\bf 304}, 10. 

\bibitem{W}
Watari, S., 2009, {\it Space Weather}, {\bf 6}, S12003.

\bibitem{Wolf}
Wolf, R., 1852, {\it Viertel. Natur. Ges. Bern}, {\bf 245}, 179. 

%\bibitem{XB}
%Xapsos, M.A., \& Burke, E.A., 2009, {\it Solar Phys.}, {\bf 257}, 363.

\bibitem{XWWL}
Xu, T., Wu, J., Wu, Z.-S., Li, Q., 2008, {\it Chinese J. Astron. Astrophys.}, {\bf 8}, 337.

\end{thebibliography}
%
% Without BibTeX 
% \begin{thebibliography}{}
% \bibitem[\protect\citeauthoryear{Author}{Year}]{key}
%   <bibliographical entry>
%
% \bibitem[\protect\citeauthoryear{}{}]{}
%   
%  
% \end{thebibliography}

\end{article} 
\end{document}